\documentclass[10pt]{article}

\def\>{\rangle}
\def\<{\langle}

\def\be{\begin{equation}}
\def\ee{\end{equation}}
\def\bq{\begin{quote}}
\def\eq{\end{quote}}
\def\ni{\noindent}
\def\text{\mbox}

\begin{document}

\title{The Laws of Physics and Cryptographic Security}

\author{Terry Rudolph}

\date{\today}

\maketitle

\begin{abstract}
This paper consists of musings that originate mainly from
conversations with other physicists, as together we've tried to
learn some cryptography, but also from conversations with a
couple of classical cryptographers. The main thrust of the paper
is an attempt to explore the ramifications for cryptographic
security of incorporating physics into our thinking at every
level. I begin by discussing two fundamental cryptographic
principles, namely that security must not rely on secrecy of the
protocol and that our local environment must be secure, from a
physical perspective. I go on to explain why \emph{by definition}
a particular cryptographic task, oblivious transfer, is
inconsistent with a belief in the validity of quantum mechanics.
More precisely, oblivious transfer defines states and operations
that do not exist in any (complex) Hilbert space. I go on to
argue the fallaciousness of a ``black box'' approach to quantum
cryptography, in which classical cryptographers just trust
physicists to provide them with secure quantum cryptographic
sub-protocols, which they then attempt to incorporate into larger
cryptographic systems. Lest quantum cryptographers begin to feel
too smug, I discuss the fact that current implementations of
quantum key distribution are only technologically secure, and not
``unconditionally'' secure as is sometimes claimed. I next
examine the concept of a secure lab from a physical perspective,
and end by making some observations about the cryptographic
significance of the (often overlooked) necessity for parties who
wish to communicate having established physical reference frames.
\end{abstract}

\vskip 0.5cm

\centerline{\textbf{\emph{Nihil est dictum, quod non est dictum
prius\footnotemark\footnotetext{By inserting this caveat at the
start I am essentially trying to avoid a whole bunch of people
being upset with me. I am writing this article off the top of my
head, with only reference [1] actually in front of me, and
reference [2] ingrained in my thick skull. Thus it should be
understood that any work which I fail to cite must be of such
quality that it has passed into the realms of ``common
knowledge''; accordingly I hereby completely disavow any claims
as to the originality of any thought expressed in this paper.
Thankfully, since I can't imagine this paper appearing anywhere
other than the quant-ph archive, I have no anxiety about anyone's
citation rates suffering.}}}}

\vskip 0.8cm

Joe Kilian began his well known STOC paper [1] on oblivious
transfer with the words \emph{Cryptographers seldom sleep well}.
Today some physicists would, rather smugly, qualify this by
inserting the word ``Classical'' in front of ``{Cryptographers}''.
As I begin writing this article at 4am on a freezing cold Viennese
morning, I can assure you that quantum cryptographers also seldom
sleep well. By the end of this article I hope you realize that
this is not due solely to jetlag.

{\center\section*{Two basic cryptographic principles}}

The field of Quantum Information has brought attention to the
fact that information theory needs to be considered in the light
of physics. In this paper I attempt to consider the ramifications
for cryptographers of thinking physically about security.

The two cryptographic principles I will begin by focusing on are
the following:
\bq
\ni\emph{Principle 1:} (Kerckhoff) The security of a cryptographic protocol
must not depend on features/details of the protocol itself being
kept secret.
\eq
\bq
\emph{Principle 2:} All parties involved in a cryptographic protocol
must operate in a secure local environment.
\eq
Since we are concerned in this paper with the role of physics in
cryptography, it is useful to rewrite these principles in a manner
that places particular emphasis on the physics:
\bq
\emph{Physicist's Version of Principle 1:} The security of a
cryptographic system must not rely on keeping secret either the
specific physical devices or the sequence of
physical/experimental operations being used to implement the
protocol.
\eq
\bq\emph{Physicist's Version of Principle 2:} All parties involved in a cryptographic
system operate in secure laboratories, and each party trusts only
in the integrity of their laboratory and not in any way on
physical events taking place in the universe external to it.\eq
\vskip0.3cm

A cryptographic \emph{task} is defined by an abstract set of
requirements. A cryptographic \emph{protocol} is designed to
achieve these requirements, and is normally presented as an
abstract series of procedures to be undertaken by the parties
involved. The abstract protocol makes reference to such notions as
``bits'', or ``qubits'' and actions such as ``choose at random…''
and so on. Whether the `bits' used are black and white pigeons or
flashes of blue and green lights is irrelevant to this abstract
formulation of the protocol. When it comes to implementing a
cryptographic protocol however, the parties involved need to
decide upon a sequence of physical operations to perform using
actual physical devices. What philosophers call a
\emph{bridge principle} - between the formal mathematics and
idealized statements/instructions of the abstract protocol and
the ``dirty experimental implementations'' - is necessarily
invoked. This is not news to a physicist: the study of physics is
founded upon the quest to uncover and elucidate the bridge
principles between mathematics and experimental procedures and
observations.

Of course there are usually many different possible physical
realizations of any given abstract cryptographic protocol, and it
is important to realize that once one particular implementation is
chosen, we are adding a link to the `chain of belief', upon which
hangs our feelings of security. No matter how much we may believe
the formal laws of mathematics, information theory or physics,
which are invoked in proving security of the abstractly defined
protocol, we need also to believe that the particular sequences
of experimental operations being performed are, in fact,
isomorphic to the abstract protocol under the bridge principle we
have chosen to believe. This observation is important beyond
merely metaphysical musings. In general, the practical realization
of the abstract protocol involves many more physical elements and
degrees of freedom than those theoretically required to implement
the protocol in its purest and simplest form. These are generally
 ignored while proving formal notions of security for the
abstract protocol. However they must be considered if we wish to
argue that a particular implementation is truly as secure as the
abstract protocol might suggest. For example, the two
polarization states of a photon might make a particularly simple
physical realization of a bit; but one must understand that a
photon is described by more degrees of freedom: its frequency,
orbital angular momentum or spatial mode for example. These
degrees of freedom are potential carriers of information too. Such
degrees of freedom are not normally innocent bystanders - for
example we might need to couple to them for transmission or
measurement purposes.

The \emph{Physicist's Version of Principle 1} is simply the
statement that, in addition to not relying on the details of the
abstract protocol being kept secret, we must not rely on keeping
secret the sets of physical devices and operations being used to
implement the protocol. Thus we assume that our adversaries know
what type of photodetectors and computers and lasers and so on we
have in our laboratories, how we intend to use them, and what
physical systems we are exchanging to affect communication.

The \emph{Physicist's Version of Principle 2} is meant to
encapsulate the fact that our feelings of security should not
rely on presumptions about systems which are not under our direct
control and observation. Ideally, we should feel secure despite
the possibility that the whole `big bad outside world' is
controlled by our adversaries, and that they have arbitrarily
large resources. That is, they are not limited technologically in
the attacks they can mount against us, they are limited only by
the laws of physics.

{\center\section*{The \emph{necessity} of incorporating physics
in cryptography}}

From the conversations I have had with a few classical
cryptographers, I will wildly extrapolate and speculate that the
majority of classical cryptographers would prefer to ignore the
much-hyped excitement about quantum information theory and its
implications for cryptography. Who, after all, wants to get
involved in learning dirty physics when we can study cryptography
through beautiful and pure mathematics? It is a sentiment I am
not altogether unsympathetic towards. A slightly more enlightened
group seem to want ``black box'' quantum cryptography. They would
like physicists to give them black box modules of quantum
cryptographic sub-protocols and routines, the absolute security
of which we will affirm (and they will, rather amazingly,
believe). They will then go off and try and incorporate these into
a larger (classical) cryptographic framework. In this section I am
going to argue that both of these perspectives are fundamentally
flawed. I'll do so by arguing that the abstract \emph{definition}
of a standard cryptographic task is incompatible with a belief in
the validity of the laws of quantum mechanics. Thus attempts to
find protocols for this task are inevitably doomed to failure. I
will then illustrate that building larger protocols out of
`promised secure' sub-protocols is a very tricky business, which
necessitates incorporating physical considerations at every step.

These days, of course, a quantum information theorist might try
to make the case for cryptographers learning some physics based
around the well-known story of quantum computation, Shor's
factoring algorithm and its obvious implications for RSA (and
many other cryptographic protocols). However this argument would
imply that cryptographers somehow thought that factorisation was
more than just technologically difficult. Protocols based on
one-way functions have always had (and will always have) security
that is fundamentally founded in a belief about the technological
capabilities of one's adversary.

A popular conception of cryptography is that it investigates how
well two co-operating parties can send secret messages. This is,
of course, only the beginning.  One of my favourite areas of
cryptography is surely one of the
simplest:\footnotemark\footnotetext{Both the voices in my head
assure me that you need at least two parties for meaningful
communication.} the study of two-party protocols. These are
protocols designed to complete cryptographic tasks, such as
coin-flipping, bit commitment or oblivious transfer, generally
between two antagonistic parties. Two-party protocols are, in
fact, some of the most problematic in classical cryptography;
there are no known secure such protocols. In [1], which has the
wonderfully ambitious title \emph{Founding Cryptography on
Oblivious Transfer}, Kilian explains:
\begin{quote}
[In a two-party protocol] both parties possess the entire
transcript of the conversation that has taken place between them.
Thus, in a protocol between A and B, A can determine exactly what
B knows about A's data. Because of this knowledge symmetry
condition, there are impossibility proofs for seemingly trivial
problems. [Classical] Cryptographic protocols ``cheat'' by
setting up situations in which A may determine exactly what B can
infer about her data, from an information theoretic point of
view, but does not know what he can easily (i.e. in probabilistic
polynomial time) infer about her data. \emph{From an information
theoretic point of view, of course, nothing has been
accomplished.} A message encrypted using a one-way permutation,
for example, might as well be sent in the clear in terms of
information theory. (emphasis added)
\end{quote}

Faced with this obstacle, cryptographers have attempted to
understand the structure of two party protocols by examining the
relationship between different tasks under assumptions that
perfectly secure protocols for certain fundamental tasks exist. A
particularly interesting such fundamental task is oblivious
transfer (OT), upon (a perfectly secure version of) which
arbitrarily secure versions of all other two-party protocols could
presumably be built. OT is designed specifically to break the
`knowledge symmetry' referred to above, and can be defined as
follows [1]:
\bq
{\it Oblivious Transfer:} Alice has a secret bit $b$. At the end
of the protocol, one of the following two events occurs, each with
probability 1/2. (i) Bob learns the value of $b$. (ii) Bob gains
no further information about the value of b (other than what Bob
knew before the protocol).

At the end of the protocol, Bob knows which of these two events
actually occurred, and Alice learns nothing.
\eq
It is useful to describe hueristically how an OT protocol would be
viewed from a physical perspective. Alice and Bob engage in
rounds of communication, which is implemented by physical systems
exchanged between them. Alice follows one of two different
sequences of actions, according to whether her secret bit is 0 or
1. These actions amount to different physical interactions, both
with apparatuses in her laboratory and with the systems that are
being exchanged. At the end of the protocol Bob reads the output
of a device in his possession. The output of the device is, with
equal likelihood, either $b$ or $\#$, and Alice cannot tell which
of the two outcomes occurs.

It is also useful to make the following observation: cryptographic
tasks do not require subjective decisions by human observers.
Thus they can be completely automated. We can imagine that Alice
and Bob have each programmed the devices in their lab to follow
some sequence of actions; the decisions that we more normally
imagine Alice and Bob themselves making, are instead made by a
Turing machine. In this way we can assume that they do not
actually look at the pieces of paper on which are written the
outcomes of measurements until the termination of the protocol.
This (possibly pedantic) assertion is an attempt to bypass the
pitfalls (and nasty emails) which arise from failing to address
the pet philosophies of everyone who has an opinion on what is
commonly known as the `quantum measurement problem'. If we believe
the laws of quantum mechanics, then the automated physical
devices that Alice and Bob must use to implement OT (or for that
matter any cryptographic protocol) will always have a description
in some (generally very large) Hilbert space.  All the actions
which take place during the protocol are described by unitary
evolution within subspaces of this Hilbert space. When Alice and
Bob finally interact with their apparatuses at the culmination of
the protocol, they perform a measurement which tells them what
the outcome of the protocol in fact was.

The definition of any cryptographic task can then be viewed as a
set of constraints on the large state of all the automated
physical systems of Alice and Bob, as well as things in the
universe external to their labs (such as communication channels
etc). In particular, we generally have specific requirements on
the statistics of outcomes which Alice and Bob can ``collapse''
this large state to, when they finally observe the output of their
devices. Thus it is natural to ask whether or not the constraints
imposed by a given cryptographic task can be fulfilled \emph{in
principle}; more precisely, whether there exists states in an
arbitrarily large complex Hilbert space, along with positive
operators corresponding to the possible outcomes of the task,
which satisfy the requirements by which the task under
consideration is defined. If there does not exist such states and
operators, I will refer to the task as being \emph{inconsistent
with a belief in quantum mechanics}, or, for conciseness, simply
as an
\emph{inconsistent task}.



To see that OT is an inconsistent task, we need to precisely
formulate the requirements OT in quantum shmantum language. The
first requirement is two states $|\psi^0_{ABU}\>$ and
$|\psi^1_{ABC}\>$, which describe completely, just prior to Bob
finding out either $b$ or $\#$, the systems of all Alice and Bobs'
apparatuses and all the systems they exchanged during the
protocol, as well as systems which lie outside their labs. The
labels $A$, $B$ and $U$ are used to identify systems lying in
Alice's lab, Bob's lab and the rest of the universe respectively.
The states are labeled by a 0 and 1, since Alice follows a
different sequence of actions if she's transferring a 0 than if
she's transferring a 1.  Its important to note that these states
must be
\emph{pure} if Principle 2 is to be followed: If my adversary has an
arbitrary powerful technology (within the laws of physics), and
they can control everything external to my lab, then I must assume
they hold a \emph{purification} of any mixed state that I hold.
Put another way: A method of ensuring that both parties describe
the complete system by a mixed state, maybe through a guaranteed
noisy channel or more generally a guaranteed decohering
environment, is equivalent to having some form of a trusted third
party - and this moves us beyond two-party protocols. The next
requirement is that there exist a three outcome POVM measurement
$\{E_0^B,E_1^B,E_{\#}^B\}$ which Bob implements at the culmination
of the protocol, on the systems of $B$, that has the properties:
$Tr(\rho_b^B E_{b}^B)=Tr(\rho_b^B E_{\#}^B)=1/2$ (Bob has equal
likelihood of obtaining $b$ or $\#$ as the protocol outcome), and
$Tr(\rho_b^B E_{\bar{b}}^B)=0$ (Bob cannot obtain the outcome to
be 1 if Alice was transferring a 0 and vice versa), where
$\rho_b^{B} = Tr_{AU}(|\psi^b_{ABU}\>\<\psi^b_{ABU}|)$. We also
require that Bob cannot estimate whether Alice was sending a 0 or
a 1, better than by implementing the three outcome measurement
and randomly guessing in the event of the $E_{\#}^B$ outcome.
This supplies an extra constraint which be written as
$D(\rho_0^{BU},\rho_1^{BU})=1/2$, where $\rho_b^{BU} =
Tr_A(|\psi^b_{ABU}\>\<\psi^b_{ABU}|)$ and $D(\cdot,\cdot)$
denotes the trace norm distance\footnotemark.
\footnotetext{$D(x,y)\equiv\frac{1}{2}Tr|x-y|$ is a measure of the
distinguishability of two density operators $x,y$.} Note that
this requirement has been explicitly formulated to allow for the
fact that Bob might have access to the rest of the universe, an
assumption that Alice must make. Finally, although Alice knows
whether the state is $|\psi^0_{ABU}\>$ or $|\psi^1_{ABU}\>$, we
require that there is no measurement she can perform which allows
her to estimate whether Bob got $b$ or $\#$ with probability
better than 1/2. Lets imagine Alice was transferring a 0. This
constraint implies that for all POVM's $\{F^{AU},I-F^{AU}\}$ we
require $\<\psi^0_{ABU}|F^{AU}\otimes E_0^B
|\psi^0_{ABU}\>=\<\psi^0_{ABU}|F^{AU}\otimes E_{\#}^B
|\psi^0_{ABU}\>$. Once again we have incorporated an assumption
that Bob must make, namely that a cheating Alice has access to the
rest of the universe.

Having formulated the requirements mathematically, we can now ask
whether they are consistent. Unfortunately we find that,
\emph{states and operators
satisfying the above requirements do not exist in any dimension
complex Hilbert space.} I leave the proof of this as an exercise
for the reader. It is important to realize that even if OT
\emph{were} a consistent task, the question would still remain as to
whether there actually exists a protocol to implement it. That is,
we would have to decide whether there existed a way of Alice and
Bob engaging in rounds of communication such that they both feel
secure that the states $|\psi_{ABU}^b\>$ are attained. The task
of secure bit commitment (discussed below) is also inconsistent
with quantum mechanics, hence we should not be surprised (with the
 benefit of hindsight!) that it has been proven that there
exists no protocol to implement it. By contrast, coin flipping
(which I will not discuss further here)
\emph{is} a consistent task, however the question remains open
as to whether there exists a protocol to implement
it\footnotemark.\footnotetext{For the sake of clarity, in this
discussion I am avoiding the issues that arise if we generalize
our definitions of tasks to allow for their implementation to
only be ``arbitrarily close to secure'' (security is achieved as
some parameter in the protocol is taken to infinity). Thus, I
more precisely should have said that \emph{ideal} coin flipping is
a consistent task, but Lo and Chau have proven that it is
impossible to find a protocol to implement it. Arbitrarily secure
coin flipping is also a consistent task, and it is actually the
question of whether a protocol exists to implement it that remains
open.}

The second point I want to raise in this section is directed at
those cryptographers who would like physicists to supply them with
secure black-box protocols.

As mentioned above, OT forms a powerful building block for
implementing secure versions of other cryptographic tasks. In
particular let us examine how an arbitrarily secure bit
commitment (BC) could presumably be founded on a secure OT
protocol. First we define bit commitment:
\bq \emph{Bit Commitment:}
Alice has a secret bit $b$. A bit commitment protocol consists of
two phases. \\
Commitment phase: Alice and Bob supplies Bob with some
token/information that depends on the value of $b$.\\
 Unveiling phase: Alice provides Bob with a further token/information from
which Bob can determine $b$.

Ideally, Alice is unable to change the value of $b$ after the
commitment phase, and Bob is unable to determine the value of $b$
before the unveiling phase.
\eq

Now lets look at a proposal (based on the one in [1], where it is
attributed to Cr\'{e}peau) for building arbitrarily secure BC on
perfectly secure OT:
\bq
\emph{BC built on OT:}\\
\ni{Commitment Phase:} Alice randomly chooses $N$ strings, each of
$M$ bits. She chooses each of the $N$ strings such that their sum
modulo 2 (their parity), is $b$. She then obliviously transfers
each bit of every string to Bob, using the perfectly secure OT protocol.\\
\ni{Unveiling Phase:} Alice announces the value of every bit.
\eq

The intuition behind the security of this BC built on perfectly
secure OT is as follows. If Bob is cheating, he needs to know the
parity of one of the N strings. For any given string, the
probability he does not obtain the outcome $\#$ is $(1/2)^M$.
Thus the probability of him being able to determine Alice's bit
$b$, is $N/2^M$. If Alice is cheating she and wants to change the
parity of a given string, she needs to announce one bit value to
be different from the one she actually transferred. She will get
away with this only if Bob obtained the outcome $\#$ for this
bit, that is, she succeeds with probability $1/2$. To change her
commitment she will need to change the parity of $N$ strings, her
probability of cheating successfully is therefore $1/2^N$. Thus
we conclude that the probability of either party cheating
successfully can be made arbitrarily small by choosing $M$ and
$N$ large enough.

Now let us imagine that we only have a \emph{partially secure}
oblivious transfer protocol. By this I mean a protocol which,
regardless of Alice's actions, she cannot increase the
probability of Bob obtaining the $\#$ outcome to greater than $p$
for some $p<1$, and, regardless of Bob's actions, he cannot
estimate $b$ with probability greater than $q$ for some $q<1$.
Could such a protocol be used to implement BC? Generalizing the
above argument, Bob's probability of successfully estimating the
bit is $Nq^M$, while Alice's probability of successfully changing
her commitment is $p^N$. These can be made arbitrarily small by
choosing $N,M$ large enough for any $p,q<1$.

It turns out that there are, in fact, partially secure OT
protocols in quantum cryptography. Here is a simple example: If
Alice is honest then she obliviously transfers $b$ by sending Bob
$|\psi_b\>$, where $|\psi_0\>=\cos\theta|0\>+\sin\theta|1\>$ and
$|\psi_1\>=\cos\theta|0\>-\sin\theta|1\>$ for some
$\theta<\pi/4$. If Bob is honest then he implements the POVM for
unambiguously discriminating these two states:
$E_0=\frac{1}{1+\cos2\theta}{\sin^2\theta\;\;\sin\theta\cos\theta
\choose
\sin\theta\cos\theta\;\;\cos^2\theta}$,
$E_1=\frac{1}{1+\cos2\theta}{\cos^2\theta\;\;\;-\sin\theta\cos\theta
\choose
-\sin\theta\cos\theta\;\;\sin^2\theta}$, $E_{\#}=I-E_0-E_1$. With
probability $1-\cos2\theta$ he obtains the correct outcome
$b=0,1$; otherwise he obtains the $\#$ outcome (indicating that
we should choose $\cos2\theta=1/2$). If Alice is cheating,
however, she can maximize the probability of Bob obtaining the
$\#$ outcome by sending him the state $|0\>$; he then obtains
$\#$ with probability $p=2\cos2\theta/(1+\cos2\theta)$.  If Bob
is cheating he can maximize his information gain about which bit
Alice is sending by implementing a Helstrom measurement; with
probability $q=(1+\sin2\theta)/2$ he would correctly identify $b$.
%

In accordance with the arguments presented above, we might
conclude that we could take this partially secure quantum OT
protocol and use it as a black box module to build arbitrarily
secure BC. However this is not the case, as is hopefully clear
from what follows:

Let us imagine imagine Alice is equally likely to want to unveil
$b=0,1$, and we are building BC on the partially secure quantum
OT protocol described above. Here is possible way for Alice to
cheat: She follows honestly the protocol for committing $b=0$,
with one small twist. The protocol specifies she must randomly
choose a whole set of even parity bit strings. She could do this
by using some random number generator, recording the even parity
strings and following the OT protocols accordingly. However she
can also take some ancillary systems that are in a quantum
superposition of every possible even parity string of $M$ bits.
She then follows the sequence of actions for obliviously
transferring $0$'s or $1$'s
\emph{controlled on} the value of a corresponding ancilla bit\footnotemark. \footnotetext{It was Dominic Mayers' crucial
insight about the ability to keep  random choices at the quantum
level which allowed him to extend the MLC thorem to proving the
impossibility of any arbitrarily secure quantum bit commitment.}
This is really the toughest part to explain to someone without
any knowledge of quantum mechanics. Suffice to say that this
leaves Alice with a large
\emph{entangled} state between the ancillary systems used for choosing
the parity strings, and the systems being sent to Bob to
implement the OT. She holds what is sometimes known as a
``purification'' of the systems being held by Bob. Its important
to realize that Alice is playing absolutely honestly at the level
of each individual OT, it is just that the random choices which
the protocol for building BC on OT specified she should make are
being performed using quantum systems instead of classical coin
tosses. Bob has no way of checking how Alice chose to generate the
random strings\footnotemark.
\footnotetext{Since it is generally believed there is no other
way of generating true randomness than quantum mechanically, it
could be argued she is playing ``more honestly''!} In actual
fact, at this stage Alice has no way of knowing for each bit
whether she has actually obliviously transferred a 0 or a 1 (she
could easily find out by measuring her ancillary systems - the
equivalent of looking at the outcome of a coin toss, but that
would destroy her ability to cheat)\footnotemark.
\footnotetext{In quantum shmantum language, Alice holds a
purification of $W_0$, where we define
 $W_0=\rho_E\otimes\rho_E\otimes(N \text{copies})$,
$W_1=\rho_O\otimes\rho_O\otimes(N \text{copies})$, and $\rho_E$
($\rho_O$) is an equal mixture of all even (odd) parity tensor
products of $|\psi_0\>\<\psi_0|,|\psi_1\>\<\psi_1|.$}

If Alice decides she wants to unveil a 0 she makes a certain
measurement on all the ancilla systems she holds, and announces
the outcomes to Bob as the even parity strings she sent. If,
however, she decides she wants to unveil a 1, then she performs a
different measurement on the ancilla systems and announces those
outcomes as the odd parity strings she sent. She is certainly
\emph{not} guaranteed of passing Bob's checks, but she succeeds
in unveiling the bit of her choice with considerably better
probability of success than one might expect from the naive
intuition discussed above. In fact her probability of successfully
unveiling the bit she wants (and not being caught cheating!) is
$(1+f^2)/2$, and I'll talk a little more about the value of $f$
below.

Bob meanwhile can cheat in a different way. Bob's goal is not to
actually determine the value of every bit that Alice obliviously
transferred. What he needs to know is whether the set of systems
\emph{as a whole} are in a state corresponding to even or odd
parity. It turns out that he can make a joint measurement on all
the systems, which is designed to distinguish only these two
possibilities. His probability of successfully identifying the
bit Alice is committing to is $(1+d)/2$.

A bit of tedious algebra shows that as we vary $M$ and $N$, we
certainly do vary the parameters $f$ and $d$ between 0 and 1.
However, for this particular pair of cheating strategies one finds
that we
\emph{always} have $f+d\ge 1$. That is, no choice of $M,N$ gives
$f\rightarrow 0, d\rightarrow 0$, as we would wish for an
arbitrarily secure protocol\footnotemark\footnotetext{In quantum
shmantum language, Alice's probability of unveiling the bit of
her choosing for this cheating strategy is given by
$(1+F(W_0,W_1)^2)/2$, where $F(\omega,\tau)\equiv
Tr|\sqrt{\omega}\sqrt{\tau}|$ is the fidelity between  two mixed
states. Bob's probability of cheating is $(1+D(W_0,W_1))/2$}. In
fact the cheating strategy described here is not the optimal one
for Alice (the one for Bob is). If we solve for their optimal
cheating strategies we do however find that $f^{\max}<1$,
$d^{\max}<1$, and thus on top of our OT protocol with partial
security we have built a BC protocol with partial security.

I should emphasize that the above argument only proves that
building BC on top of partially secure OT using this one
particular suggestion does not result in an arbitrarily secure
protocol. The question of exactly how secure BC can actually be
made is still open, but it is known that it can never be made
arbitrarily secure from the work of Mayers, Lo and Chau.
Reasonably tight bounds on the best possible security for BC are
discussed in [2]. The crucial point I want to make with this
example however, is that
\emph{we must be very careful when constructing
larger cryptographic systems based on (even well understood)
sub-systems.}

Now I know that there are a whole lot of physicists who (if they
actually read this paper) will be ready to jump down my throat
with arguments against an implicit assumption I've been making,
that quantum mechanics is somehow applicable to the physical
systems of any protocol regardless of their size, shape, color or
ethnicity. And this raises exactly the third essential point I'm
trying to make in this section. If someone truly believes this and
designs a protocol based around it, then their faith in the
security of a cryptographic scheme is still founded in their
(supposed) understanding of physics; moreover it is founded in an
belief which is not, by my reckoning, the accepted norm amongst
physicists. A less dogmatic argument would be to assert that
perhaps some cheating strategies are technologically infeasible;
for example they might require Alice to maintain large amounts of
entanglement. This may well be true, but then we are still back
to admitting we have a sense of security which is founded on a
belief about the technological capabilities of our adversaries.

{\center\section*{The \emph{desirability} of incorporating
physics in cryptography}}

In this section I will briefly discuss reasons I think two-party
quantum cryptography should be of interest to all cryptographers.

Firstly there is an analogy with one-way functions\footnotemark
that non-orthogonal quantum states provide. \footnotetext{One way
functions in classical cryptography are functions for which
$f(x)=y$ is easy to compute, but finding $x=f^{-1}(y)$ is
computationally difficult.} As Kilian pointed out, because the
difficulty of inverting one-way functions is only computational,
founding cryptographic protocols on 1-way functions does not
provide `information-theoretic' security. One-way functions are
generically used in scenarios where Alice and Bob have agreed on
the particular function, Alice sends Bob one (or some) outputs
$\{y_i\}$ of the function and she feels confident that Bob has
not the computational resources to find the corresponding
$\{x_i\}$. We can find a quantum analogy as follows: Alice and Bob
agree on a particular set of states, Alice sends Bob one (or
some) states $\{\psi_i\}$. If the states are non-orthogonal then
Alice knows that the laws of physics prevent Bob determining
exactly which states have been sent. Unfortunately they also
prevent Bob verifying with absolute certainty what state Alice
has sent. But I suspect that the fundamental tradeoff between
these two degrees of certainty lies at the heart of determining
the absolute limits to the security of all two-party protocols.

Secondly, quantum cryptography provides an avenue for the
breaking of  ``knowledge symmetry''. As discussed above, OT
protocols with bounded degrees of security are certainly possible
in quantum cryptography, and these break the knowledge symmetry
referred to by Kilian. This also has implications for game theory,
the analysis of two player games shares a lot in common with the
analysis of two-party cryptography. Unfortunately no arbitrarily
secure two-party quantum cryptographic protocols are known, in
fact the only such protocol we really know a lot about in the
context of quantum cryptography is BC. Interestingly, if the two
parties share a trusted resource of entangled states, arbitrarily
secure BC is still impossible, and in fact the security bounds
remain unaffected. However, coin-flipping with trusted shared
prior entanglement is trivial. There is clearly an intricate
hierarchy of these protocols.

Finally, quantum protocols naturally display
\emph{cheat sensitivity.} When we use a classical one-way function,
we implicitly assume that our adversary is trying their best to
invert it. They have nothing to lose by doing so. However when
communication is being implemented by transfer of quantum states,
the actions of cheating parties generically lead to disturbances
of the states and this in turn leads to situations wherein one can
detect whether one's adversary is trying to cheat or not. Since we
must presume that the two-parties engaged in any two-party
protocol are not so adversarial that they have no motivation to
communicate at all, it is not unreasonable to start analyzing
situations in which the parties assign certain \emph{costs} to
being caught cheating. Furthermore, we generally find that when
one party cheats in a quantum protocol it results in a lowering
of their ability to detect whether the other party is also
cheating. There are all sorts of interesting scenarios and
tradeoffs along these lines which are almost completely
unexplored;  they can surely fail to interest only the most
apathetic and unimaginative persons in the cryptographic
community.

{\center\section*{Different conceptions of security}}

Up until now I have been throwing the word security around a
little carelessly. In this section I am going to try to
differentiate more precisely the three concepts of security that
I carry around in my mind on a day to day basis.

The first concept is \emph{Technological Security}, or T-security.
This is most common type of security employed in our day to day
cryptography.  T-security is of course the reason that there is a
seemingly endless game of back and forth between cryptographers
and cryptanalysts, and this keeps a lot of people employed. There
is no need to worry about their job security yet - later I will
argue that `almost all' security is technological, despite what
quantum cryptographers might try and assert to the contrary.

Within the category of T-security one can identify two themes.
Consider first RSA public key cryptography, the security of which
is based on the difficulty of factorising large numbers. If I use
the RSA algorithm I do so in the belief that my adversaries do
not have the technological capability to factorise the large
numbers required to compromise my messages - factorisation is not
in principle impossible. Moreover, I must accept that if my
adversary wishes to, they can keep trying for the next 20 years
to decode my message, and perhaps a quantum computer will be
built before a decoding of the message has lost its value. Now
consider quantum key distribution (QKD). In the next section I
will argue that most, if not all, current implementations of
quantum key distribution are also only T-secure. However there
seems to be an intrinsic difference with the T-security of RSA.
If the eavesdropper does not have the technological capability to
render the QKD insecure \emph{during the implementation of the
protocol}, then it would seem she has lost forever the chance to
do so.

The second concept of security is \emph{Information Theoretic
Security} (IT-security). This is the level of security most
desired by classical cryptographers - it is security that does
not rest on complexity-theoretic assumptions such as the
difficulty of factoring, and is therefore viewed by classical
cryptographers as being superior to T-security. However, as I
have discussed above, IT-security might not actually be as
attainable as one might hope. An important question is whether in
fact any protocols are IT-secure? From one perspective we could
look at Quantum Information Theory (QIT) as just being a larger
theory of information which incorporates Classical Information
Theory. We can then argue that as long as a protocol is
QIT-secure, it really and truly \emph{is} secure, and the problem
with the standard IT-security is that it was actually only
``Classical IT''-secure. While a part of me wants to believe this
(if only because I work in quantum information theory!), I'm not
sure I can make an unassailable argument for the ultimate
ubiquity and universal applicability of the principles of QIT.

As such, the third concept of security I like to carry around in
my head is \emph{Physical Security} (P-security), which is
security premised upon the truth of the laws of physics as we
currently understand them. This type of security is clearly still
a ``degree of belief'' - it is not unknown for laws of physics to
be modified as we deepen our understanding of the universe.
Moreover; we do not have one universal theory of physics
applicable to all energy scales. Thus when confusion could arise,
we should in fact qualify which particular physical theory we are
basing our cryptographic protocol around. The most obvious
candidates are non-relativistic quantum mechanics, quantum field
theory, special relativity\footnotemark and general relativity.

\footnotetext{In fact Kent has devised a P-secure two-party protocol
akin to bit commitment based on special relativity. I say ``akin''
because the protocol requires an extra `holding phase' during
which communication still takes place, as well as guarantees
about the spatial proximity or otherwise of certain communicating
parties. These constraints are difficult to fit into a standard
cryptographic framework, but the protocol is certainly
nonetheless interesting.}

The multiplicity of physical theories upon which P-secure
protocols could be based might make a non-physicist distinctly
uncomfortable. For this reason, cryptographers attempting to
construct P-secure protocols should do so by trying to use those
features of the physical theory which are expected to survive in
the ``\"{u}ber-theory'' that most of us believe will one day be
discovered. Ideally, P-secure protocols would manifestly rely only
on some general physical principles - such as conservation laws,
no-signaling or perhaps, for quantum mechanical theories,
linearity. It should be emphasized that even if we discover an
\"{u}ber-theory of physics which withstands repeated experimental
test across all accessible energies, security based on this
theory is still founded on a degree of belief in it. Nature will
never hand us out a certificate congratulating us on having, in
fact, discovered everything that is to be discovered. It is
interesting to speculate that experimental physicists would
primarily place their faith in an \"{u}ber-theory based upon it
passing multiple experimental tests, whereas the foundation of a
theoretical physicist's faith would presumably be the
mathematical beauty and elegance of the theory. Faith, trust or
belief in something is ultimately a personal and often very
subjective decision. The majority of people who use cryptography
are in fact placing their faith in cryptographers; they have
neither attempted to design an algorithm for factorization nor to
test the laws of quantum mechanics.

{\center\section*{Why quantum cryptographers should not be too
smug}}

As promised near the beginning of the previous section, in this
section I am going to try and argue that most, if not all, current
experiments implementing QKD are actually only T-secure. This
particular attack relies heavily on the ideas received from
multiple communications with Richard Gill and Gregor Weihs. Let
me say emphatically that I am not asserting the insecurity of QKD
in an idealized, abstract protocol, but rather I'm using one
particular (and presumably not optimal) technological attack on
\emph{current} implementations of QKD. Essentially I want to raise
the question about where, in our metaphorical chain of belief, we
are really placing our faith when we use QKD?

I will use the example of entanglement based QKD, due originally
to Ekert, although as C\v{a}slav Brukner pointed out to me the
attack described works equally well against the current
implementations of BB84 or B92 QKD. Entanglement based QKD gives
quantum cryptographers a nice, warm fuzzy feeling inside, because
it exploits what is (unfortunately) commonly termed ``quantum
nonlocality''. More precisely, it exploits the inability of any
set of local hidden variables to reproduce the correlations in
data recorded at two spacelike separated locations. Ideally the
QKD works because even an eavesdropper with an arbitrarily
powerful technology (i.e. constrained only by the laws of quantum
mechanics) cannot obtain a significant amount of information
about the key without degrading these correlations.

Let me give now a slightly idealized description of most current
implementations of QKD. Much of my understanding of such things
is due to Thomas Jennewein who patiently tolerates my questions -
however he should not be held responsible for inaccuracies! Alice
uses a parametric downconverter to produce entangled pairs of
photons\footnotemark.\footnotetext{The photon source could in
fact be located in-between Alice and Bob and therefore controlled
by Eve, however this unnecessary for Eve to be able to mount the
attack I'm describing here.} One of the output beam of photons is
sent to Bob (and it is these photons which Eve may intercept and
interact with). The other beam of photons is retained by Alice.
Alice sends her beam through a polarizer, which is rapidly and
randomly set to one of a choice of several rotation angles, and a
measurement of the photon's polarization is then made. We'll
assume here that Alice's photon detectors are 100\% efficient.
Alice records the time of measurement of any photon using an
atomic clock. The rate of switching polarizer settings is fast
compared to the photon production rate of the source - most of
the time no photons go through any given setting. Meanwhile Bob
is doing the same thing with the beam of photons which has been
sent to him.

After a suitable number of photons have been detected, Alice and
Bob communicate publicly. In the idealized protocol they would
announce which setting they used for each pair of photons, the
measured polarization outcomes of the subset of photons for which
they chose different polarizer settings would then also be
announced, and sufficient violation of the Bell inequality for
this subset would indicate that Eve had not gained enough
information to compromise the security of the QKD.

In reality however, Alice and Bob must also announce their photon
arrival times, and only consider photons which arrived in
coincidence. This is because photons get lost: The pairs are not
emitted perfectly in opposite directions, the quantum channel
(optical fiber/free space) is lossy and so on. Photons which do
not arrive in coincidence are not part of an entangled pair and
therefore not useful for the QKD. This allows Eve the following
attack: Eve replaces the lossy channel with one of much lower
loss rate. She then acts just like Bob - she makes random choices
of polarizer settings and records the outcomes/detection times of
measurements on the photons that were meant for Bob. She then
constructs a ``demonic'' photon. This is a photon (or wavepacket
of photons) which is programmed to go along the channel into Bob's
lab, but to only arrive in coincidence with Alice's photon if the
setting on Bob's polarizer is the same as the setting Eve
randomly chose. If it does arrive in coincidence, then it is
programmed to give the measurement outcome that Eve obtained. If
Eve and Bobs'  settings are different, the photon aborts its
mission (e.g. slows down/speeds up) in order to avoid arriving in
coincidence. Thus for all photons which Alice and Bob accept as
having arrived in coincidence, Eve actually knows the outcome Bob
obtained.

Now I know that the same people who didn't like my implicit use of
macroscopic entanglement earlier on are
\emph{really} going to hate the usage of demonic photons. How,
after all, do I propose to construct such a photon. In fact the
question is irrelevant - what is important is whether the
construction of such a photon is forbidden by the laws of quantum
mechanics. We want our security to be founded
\emph{only} on our belief in those laws, and perhaps more
specifically on the aforementioned warm fuzzy feeling we get about
Bell's inequality. Note that here the Bell inequality would still
be violated, even though the demonic photon is programmed with a
local hidden variable\footnotemark.\footnotetext{It should be
pointed out that this attack is really just an exotic
exploitation of the well known ``detector loophole'', and
therefore not of fundamental philosophical significance to Bell
inequality type experiments.} It seems to me we have reverted back
to T-security: can proponents of current experimental
implementations categorically assure their users that no demonic
photons are present? Since no-one to date has formulated the
minimal amount of complexity required for construction of such
demonic photons, I doubt it. In fact the demonic photon does not
need to be extremely sophisticated: If the two polarizer settings
in Bobs lab have slightly different optical properties
(birefringence, or dispersion characteristics say) then Eve may
be able to utilize this to create quite unintelligent but
effective demons (recall that by the \emph{Physicists Version of
Principle 1} she knows all the equipment that Bob is using).

However; the astute reader will be able to argue against the
relevance of above attack on QKD with a much better argument than
``well it seems unlikely''. Essentially Eve's attack can be
viewed as violating the \emph{Physicist's Version of Principle
2.} After all, who can claim to feel safe and secure in a
laboratory with photonic demons floating around it? More
seriously, it is important to realize that the information we
receive down a quantum channel is actually physical stuff. Proofs
of security of abstract protocols generally implicitly assume
that what comes into a lab through such a channel contains states
with support only in a certain specified subspace of the full
Hilbert space. We might say that the channel is presumed to be
exorcised. However such exorcism is not in general trivially
accomplished\footnotemark
\footnotetext{I think even the bible of QIT
(Nielsen and Chuang's textbook) does not provide us with an
operational definition of the measurement operator corresponding
to projection onto a non-demonic set of states - and if you can't
find mention of demons in a bible, where else are you going to
find them?}, since we generally rely on the many unspecified (in
the abstract protocol) additional degrees of freedom of these
systems to aid in their actual transmission or to couple to with
measurement devices.

{\center\section*{Is the \emph{Physicist's Version of Principle
2} actually reasonable?}}

In this final section I am going to outline some thoughts about
the physical limitations on having ``totally secure labs''. Let
me start with the most extreme interpretation of Principle 2 that
we can have: namely that we feel secure only about the actions
and events taking place in our lab, and that all the outside
world is potentially controlled by our adversary. To quote
Descartes:
\bq
I will suppose that...some malicious demon of the utmost power and
cunning has employed all his energies in order to deceive me. I
shall think that the sky, the air, the earth, colours, shapes,
sounds and \emph{all external things} are merely the delusions of
dreams which he has devised to ensnare my judgement. (emphasis in
original)\footnotemark
\eq
\footnotetext{I admit it - I do in fact have more than just [1]
in front of me - for when I get bored I also have \emph{The Great
Philosophers} edited by Ray Monk and Frederic Raphael. Note that
Descartes' demon is very powerful and so presumably much higher up
in the underworld hierarchy than the photonic demon of the
previous section.} Is it possible to have a lab, the security of
which is guaranteed by the laws of physics? Such a lab must be
secure against both active external attacks (impenetrable) and
passive outside monitoring (emanation free). Impenetrability must
be guaranteed for two types of penetration attacks: (i) Attacks
which involve penetration and subsequent gathering of information
by the adversary from the penetrating agents (e.g X-rays) and
(ii) Attacks in which the penetrating agents attempt to modify
the physical conditions/devices inside the lab. Such penetration
attacks can of course be countered by the person in the lab
trying to detect the penetration, but then we are definitely back
to a technological level of security again; we would be trusting
that we are as smart and advanced as the persons attempting the
penetration. The second type of penetration attack particularly
troubles me. As an extreme example, how can I be sure that my
adversary is not able to wave a large mass outside my lab in such
a way as to distort the spacetime inside my lab to their
advantage? In the limit of an infinitesimally small lab this is
not a problem, but I'm not infinitesimally small!

 An emanation free lab would not be very comfortable, since
 information processing at absolute zero is unlikely to be much
 fun - although it is not impossible. Completely emanation free
 labs are presumably unnecessary - for example, we
would not in general care if our adversaries could only measure
emanations that inform them as to our temperature. However; if the
whole lab is at thermal equilibrium, not much information
processing will be going on inside it anyway. (I could try and
elucidate on a tenuous link here to Maxwell's demon, however two
demons in any one paper should be enough for anybody.) Thus a
natural physical question, to which I do not have an answer, is
whether we can construct (hypothetical idealized) labs for which
all emanations are guaranteed to leak only an exponentially small
amount of useful information to the outside world.

The impenetrability/no emanation conditions are presumably
satisfied by a `black hole lab', and perhaps this is the only way
that they can be. Unfortunately being inside a black hole lab
will present significant hurdles to communication. In fact there
seems to be an intrinsically physical aspect to communication that
is often overlooked in quantum cryptography. If two parties are
communicating classically, it is generally presumed that they
both can distinguish a 1 from a 0, and in the majority of
circumstances this is trivial. However when we progress to
exchanging very small systems for communication, things are
considerably more subtle. Consider first using different
two-level quantum systems for
\emph{classical} communication. Exchanging two level atoms would present no
problems for classical communication - the ground and excited
states require no specialized reference frame to identify. For
degenerate systems however things are more problematic. The ``spin
up'' or ``spin down'' states of spin-1/2 systems can only be used
as classical bits if the two parties have agreed upon a reference
frame, in this case which direction is up. The setting up of such
a spatial reference frame is a physical process, requiring either
the exchange of physical systems or agreement about external
reference points (such as a particular fixed star to define a
spatial direction say, which then presumes proximity of the
parties). Using the horizontal and vertical polarizations of a
photon is in fact a little more subtle. The polarization lies in
the plane defined with respect to the photon's direction of
propagation. Thus a single `external fixed star' method of
establishing a reference frame will not work unless the two
communicating parties also know the orientation of their labs
with respect to each other - in which case the reference frame is
presumably trivially fixed. Of course, in these latter two cases,
appropriate reference frames can be set up by exchanging a large
number of photons or spin-1/2 particles, communicating
classically by other means, (using the two-level atoms say) and
then playing around with the relevant polarizers/Stern-Gerlach
apparatuses until a reference frame is fixed. The need to
communicate classically by other means could possibly be avoided
if some standard protocol was already in place; however, it is
clear that even at the simple level of exchanging
\emph{classical} bits with these objects something extra is
required. And remember, we need to feel  sure that the process of
establishing this reference frame does not allow our adversary
some sophisticated cheating strategies...

Once we want to exchange quantum systems in coherent
superpositions then even the two-level atoms are no longer a help
- the relative phase between the ground and excited states must be
fixed using some form of time standard. Now it is unreasonable to
expect (in fact its quantum mechanically impossible) any
reference frame to remain fixed for an arbitrary amount of time.
This leads to the following problem. If two parties are engaged
in quantum communication they will eventually be faced with one
of these two choices: (i) They must agree to use some external
reference points to re-calibrate their reference frame, or (ii)
they must exchange some physical systems to perform the
re-calibration. Either possibility is not ideal from a
cryptographic standpoint. For example, under (i) I might be
forced to accept that my adversaries do not have the ability to
move around the fixed stars (or at least to install faux stars
between my lab and the real stars). As such I am trusting in the
security of something external to my lab. Under (ii) I might use a
reference frame which is completely determined by my adversaries,
this too leads to obvious potential problems. We might take the
view that shared physical systems that establish reference frames
are in themselves an information theoretic resource.

%

{\center\section*{InConclusions}}

 I have tried to explore just a
tiny subset of the issues that arise when we approach
cryptography by incorporating physics into our thinking at every
level. Of course I do not anticipate the day the NSA employs more
physicists than mathematicians and computer scientists arriving
any time soon. The issues discussed here are, for the moment,
completely irrelevant - there are much weaker links in present
cryptographic systems than those raised here. It seems inevitable
to me however, that discussion about the `ultimate' limits to
security will always lie in the realm of physics. We are physical
creatures, `security' has no existence in Plato's universe and
therefore cannot be `proven'; security is a subjective feeling
based on a particular individual's beliefs.

{\center\subsection*{Acknowledgements}} In addition to the people
mentioned explicitly in the text, thanks to: Steve Bartlett,
Chris Fuchs, Lov Grover, Paul Kwiat, Barry Sanders, Peter Turner
and Anton Zeilinger for many interesting conversations. To Mark
and Paul and the guy I met on the New Jersey Transit train; while
I find your viewpoints completely untenable and in the main
absurd, you certainly deserve acknowledgement for stimulating me
to try and argue one physicist's side.  Thanks to Christian
Kollmitzer for restoring some of my faith in classical
cryptographers; I hope that I have explained why I am troubled by
``black box'' quantum cryptography. Finally many thanks to Rob
Spekkens, with whom I have had so much interaction about
cryptography and physics that I can no longer differentiate
between my thoughts and his. As such he should be given the
credit for anything that is worthwhile in this article. This
research is supported (financially but not necessarily
philosophically) by the NSA \& ARO under contract no.
DAAG55-98-C-0040.

\end{document}